\documentclass[]{JFM-FLM_Au}

\usepackage{bm}

\newcommand{\overbar}[1]{\mkern 1.5mu\overline{\mkern-1.5mu#1\mkern-1.5mu}\mkern 1.5mu}

\usepackage[svgnames]{xcolor}
\hypersetup{colorlinks=true, citecolor=NavyBlue, urlcolor=NavyBlue, linkcolor=NavyBlue}

\lefttitle{X.M. de Wit}
\righttitle{Journal of Fluid Mechanics}

\title{Transition to inverse cascade in turbulent rotating convection in absence of the large-scale vortex}

\author{Xander M. de Wit}

\affiliation{
Fluids and Flows group, Department of Applied Physics and J. M. Burgers Centre for Fluid Dynamics, Eindhoven University of Technology, P.O. Box 513, 5600 MB Eindhoven, Netherlands
}

\corresau{Xander M. de Wit, \email{x.m.d.wit@tue.nl}}

\begin{document}
\maketitle

\begin{abstract}
Turbulent convection under strong rotation can develop an inverse cascade of kinetic energy from smaller to larger scales. In the absence of an effective dissipation mechanism at the large scales, this leads to the pile-up of kinetic energy at the largest available scale, yielding a system-wide large-scale vortex (LSV). Earlier works have shown that the transition into this state is abrupt and discontinuous. Here, we study the transition to the inverse cascade at Ekman number $\textrm{Ek}=10^{-4}$ and using stress-free boundary conditions, in the case where the inverse energy flux is dissipated before it reaches the system scale, suppressing the LSV formation. We demonstrate how this can be achieved in direct numerical simulations by using an adapted form of hypoviscosity on the horizontal manifold. We find that in the absence of the LSV, the transition to the inverse cascade becomes continuous. This shows that it is the interaction between the LSV and the background turbulence that is responsible for the earlier observed discontinuity. We furthermore show that the inverse cascade in absence of the LSV has a more local signature compared to the case with LSV.
\end{abstract}

\begin{keywords}
turbulent transition, inverse energy cascade, rotating convection
\end{keywords}

\vspace{4mm}

\section{Introduction} \label{sec:intro} 
Convection is the main source of fluid motion in many of the natural flows that we see around us, from the liquid metal in the Earth's core, to the surface of the Sun and the planetary gas giants \citep{Aurnou2015,Cheng2015,Miesch2000,Evonuk2007,Heimpel2007}. The large scale nature of these geophysical and astrophysical flows, combined with the vigorous force of convection, results in very strong fluid turbulence. From the fundamental point of view, turbulence is often studied in idealized models where the driving mechanism is a well-controlled body force that is typically confined to a limited set of length and time scales. In convectively driven turbulence, by contrast, the buoyant forcing self-organizes in response to the flow, resulting in a broad-band spectrum of spatio-temporal forcing scales, giving rise to a rich dynamics of many different flow regimes \citep{Ahlers2009}. It is thus important to investigate how physical mechanisms observed in idealized turbulent systems translate to the natural flow system of convective turbulence. 

Many of the convective flows that are encountered in the geo- and astrophysical settings are also subject to strong rotation, typically due to rotation of the celestial body itself, which can dramatically alter the flow \citep{Ecke2023}. When rotation is strong, it suppresses variations of the velocity along the rotation axis, promoting quasi-two-dimensional (quasi-2D) flow organization. As a result of the quasi-two-dimensionalisation, the flow can develop a partial inverse cascade of kinetic energy, where a fraction of the kinetic energy is transferred from smaller length scales to larger scales, opposite from the direct cascade in regular 3D turbulence. In absence of an effective dissipation mechanism at the large scales, this energy will pile up at the largest available scale. This condensate of kinetic energy at the largest scales manifests itself in the form of a strong system-wide vortex, which in rotating convection has been coined the large-scale vortex or LSV \citep{Favier2014,Guervilly2014,Julien2012,Rubio2014}. Recently, it has been shown that the transition from the forward cascading 3D-like state to the inverse cascading quasi-2D LSV state in rotating convection is, remarkably, an abrupt and discontinuous transition \citep{Favier2019,DeWit2022}. In particular, the `high-thermal-forcing' transition, where the rotational constraint of the flow breaks down to due finite rotation effects, was shown to be strongly hysteretic. This is in close similarity with the condensate transition observed in other quasi-2D flow systems, such as thin-layer turbulence \citep{VanKan2019,DeWit2022b} and forced rotating turbulence \citep{Yokoyama2017}, see also \citet{Alexakis2018}, \citet{Alexakis2023} and \citet{vanKan2024} for recent reviews.

In some cases, however, the inverse cascade can be affected by a dissipation mechanism at large scales. If such a mechanism is sufficiently strong, it can dissipate the inverse energy flux entirely before it reaches the largest system scale and thereby suppress the formation of the large-scale condensate. A prime example is boundary friction, which has been shown to be able to partially or fully suppress the formation of the LSV in rotating convection \citep{kunnen2016transition,julien2016nonlinear,plumley2016effects,Guzman2020}. As system size increases, the larger scale separation between injection scales and the system scales allows a larger range for such a dissipation mechanism to be active, so that even an ostensibly small dissipation mechanism can eventually dissipate the inverse flux entirely before it reaches the system scale.
In direct numerical simulations (DNS) this scenario is typically studied by applying a hypoviscosity: a dissipation mechanism that is artificially restricted to the largest scales, in order to mimic a large degree of scale separation. Earlier works in other quasi-2D turbulent flow systems have revealed that such large-scale friction can drastically alter the inverse cascade transition \citep{Seshasayanan2014,Benavides2017,Pestana2019,VanKan2020}, where in particular, the discontinuity in the transition can be partially or entirely suppressed when the large-scale condensate is absent. In this work, we study the inverse cascade transition in the natural flow system of rotating convection under the influence of a large-scale friction that prevents the formation of the LSV. This allows us to disentangle the effect of the LSV from the dynamics of the underlying turbulence to understand its influence on the inverse cascade transition.

In section \ref{sec:numerical} we lay out the numerical methods used in this work, focusing on how one can practically achieve an effective hypoviscosity in a doubly periodic system. Section \ref{sec:hypo_example} shows a typical example case of rotating convection under the influence of this large-scale friction. In section \ref{sec:transition} we discuss how the inverse cascade transition is altered in the presence of large-scale friction and what that teaches us about the influence of the LSV on the nature of this transition. Finally, section \ref{sec:conclusion} provides the conclusions and outlook of this work.

\section{Numerical methods} \label{sec:numerical} 
We study the canonical model system of rotating Rayleigh-B\'{e}nard convection, where the fluid is contained between a hot bottom and cold top boundary with a temperature difference $\Delta T$, driving the fluid motion. The flow is simultaneously subject to a background rotation $\Omega$, aligned with the gravitational axis, leading to a Coriolis force. The system is described by three dimensionless numbers: the Rayleigh number $\textrm{Ra}$, quantifying the strength of the buoyant forcing, the Ekman number $\textrm{Ek}$, capturing the (inverse) strength of rotation, and finally the Prandtl number $\textrm{Pr}$, describing the diffusive properties of the fluid, given by
\begin{equation}
    \textrm{Ra}\equiv\frac{g\gamma\Delta T H^3}{\nu\kappa}, \qquad \textrm{Ek}\equiv\frac{\nu}{2\Omega H^2}, \qquad \textrm{Pr}\equiv\frac{\nu}{\kappa}.
\end{equation}
Here, $g$ denotes the gravitational acceleration, $H$ is the domain height, while the fluid properties $\gamma$, $\nu$ and $\kappa$ represent the thermal expansion coefficient, kinematic viscosity and thermal diffusivity of the fluid, respectively.

We employ the finite difference code developed by \citet{Verzicco1996} (see also \citet{Ostilla-Monico2015}) to perform DNS of the full Navier-Stokes and heat equations in Oberbeck-Boussinesq approximation. The domain is a $L \times L \times H$ cuboid with Cartesian basis $(\bm{e}_x, \bm{e}_y, \bm{e}_z)$ that is horizontally periodic and has stress-free, constant temperature top and bottom boundaries. The full governing equations evolving the velocity field $\bm{u}$ and temperature field $T$ in time $t$ are given in convective units by

\begin{subequations}\begin{align}
    \frac{\partial\bm{u}}{\partial t} + \left(\bm{u}\cdot\bm{\nabla}\right)\bm{u} + \frac{\left(\textrm{Pr}/\textrm{Ra}\right)^{1/2}}{\textrm{Ek}}\bm{e}_z\times\bm{u} &= -\bm{\nabla}p + T\bm{e}_z + \left(\frac{\textrm{Pr}}{\textrm{Ra}}\right)^{1/2} \Delta \bm{u} - \underbrace{ \nu_\alpha \Delta_h^{-\alpha} \bm{u}_h}_{\substack{\text{adapted}\\\text{hypoviscosity}}}, \label{eq:NS_momentum}\\
    \frac{\partial T}{\partial t} + \left(\bm{u}\cdot\bm{\nabla}\right)T &= \frac{1}{\left(\textrm{Ra}\textrm{Pr}\right)^{1/2}}\Delta T, \label{eq:NS_temp}\\
    \bm{\nabla}\cdot\bm{u} &= 0. \label{eq:NS_cont}
\end{align}\end{subequations}
Notice that in the last term of equation~\eqref{eq:NS_momentum} we have included the adapted hypoviscosity. This is the large-scale friction term that takes care of dissipating the inverse energy flux. In a doubly periodic domain like in the (rotating) convection system at hand, we need to be somewhat careful how we can define an effective hypoviscosity, as we cannot simply use the isotropic inverse Laplacian that is commonly applied in triply periodic domains. Instead, motivated by the aim of dissipating the inverse cascade that resides in the 2D horizontal manifold, we apply a 2D variant of the hypoviscosity that acts on the horizontal velocity and the horizontal inverse Laplacian, applied across the full height of the domain. This is conveniently computed from a 2D Fourier transform in the horizontal periodic directions $\mathcal{F}_h\{...\}$, taking the horizontal velocity $\bm{u}_h\equiv u_x \bm{e}_x+u_y\bm{e}_y$ from the real space $\bm{u}_h(x,y,z,t)$ to the Fourier space $\hat{\bm{u}}_h(k_x,k_y,z,t)$. Then, we can compute the hypoviscous term as
\begin{equation}
    \nu_\alpha \Delta_h^{-\alpha} \bm{u}_h \equiv \nu_\alpha \mathcal{F}_h^{-1} \left\{\frac{1}{\left(k_x^2+k_y^2\right)^\alpha} \mathcal{F}_h \left\{ \bm{u}_h \right\} \right\}.
\end{equation}
In practice, we need to avoid the singularity at $k_x=k_y=0$, so we approximate
\begin{equation}
    \nu_\alpha \Delta_h^{-\alpha} \bm{u}_h \approx \nu_\alpha \mathcal{F}_h^{-1} \left\{\frac{1}{\textrm{max}\left((2\pi/L)^2,k_x^2+k_y^2\right)^\alpha} \mathcal{F}_h \left\{ \bm{u}_h \right\} \right\},
\end{equation}
clipping the horizontal wavenumber from below by the smallest non-zero wavenumber $2\pi/L$.

A different approach for suppression of the LSV in rotating convection is put forward by \citet{Oliver2023}, based on a modification of the non-linear term. In appendix \ref{appB}, we compare the different approaches for suppression of the LSV for an exemplary case.

We consider a fluid with $\textrm{Pr}=1$, rotating at $\textrm{Ek}=10^{-4}$ and vary $\textrm{Ra}$ over the range in which the flow is known to transition into and out of the inverse cascade regime, in agreement with the parameter range considered in \citet{DeWit2022} for the case with LSV (without hypoviscosity). See table~\ref{tab:input} for the full set of input parameters considered in this work. The power $\alpha$ and strength $\nu_\alpha$ of the hypoviscosity are tuned in such a way that the full inverse energy flux is dissipated before it reaches the system scale, while simultaneously ensuring that its action remains mostly restricted to the largest scales (see also next section). We consider three series of runs with different horizontal domain sizes. The aspect ratio of the first series is chosen in agreement with \citet{DeWit2022}, employing $L=10\mathcal{L}_c$, with $\mathcal{L}_c \equiv 4.8\textrm{Ek}^{1/3}H$, the most unstable wavelength at onset of convection \citep{Chandrasekhar1961}. The second and third series of runs progressively double the horizontal extent in order to increase the scale separation between the system scale and the scales that dominate the energy injection. The hypoviscosity parameters are kept constant throughout each series of runs. The grid is uniform in the horizontal directions, but non-uniform in the vertical direction, narrowing the grid close to the top and bottom to accurately resolve the boundary layers. The resolution $N_x \times N_y \times N_z$ is chosen with respect to the Kolmogorov scale $\eta$, ensuring that the grid spacing never exceeds $2\eta$, which is well below the threshold of $4\eta$ that is obtained for this code in \citet{Verzicco2003}.

\begin{table}
\caption{\label{tab:input}Input parameters that are used for the simulations in this work.}
\vspace{3mm}
\makebox[\textwidth][c]{
\begin{tabular}{lcccccccc}
\toprule
&\#runs&\textrm{Ra}&\textrm{Pr}&\textrm{Ek}&$\nu_\alpha$&$\alpha$&$L/H$&$N_x\times N_y\times N_z$\\
\midrule
With hypoviscosity I& 22 &$[2\times10^{6} - 5\times10^7 ]$& 1 & $10^{-4}$ & $1.5\times10^{2}$ & 4 & 2.235 & $257\times257\times129$\\
With hypoviscosity II& 22 &$[2\times10^{6} - 5\times10^7 ]$& 1 & $10^{-4}$ & $3.8 \times 10^{-2}$ & 1 & 4.470 & $513\times513\times129$\\
With hypoviscosity III& 22 &$[2\times10^{6} - 5\times10^7 ]$& 1 & $10^{-4}$ & $1.1 \times 10^{-2}$ & 0.75 & 8.940 & $1025\times1025\times129$\\
Without hypoviscosity$^*$& 46 &$[2\times10^{6} - 5\times10^7 ]$& 1 & $10^{-4}$ & N/A & N/A & 2.235 & $257\times257\times129$\\
\bottomrule
\multicolumn{4}{l}{$^*$covered in \citet{DeWit2022}\rule{0pt}{8pt}}
\vspace{3mm}
\end{tabular}
}
\end{table}

Simulations are initialized with zero velocity and a linear temperature profile connecting the top and bottom boundaries, superposed by a tiny statistical perturbation to initiate convection. Statistics are then collected after 1000 convective time units when a statistically steady state is reached at all scales.

\section{Rotating convection with large-scale friction} \label{sec:hypo_example} 

Analogous to many other quasi-2D flow systems, rotating convection can give rise to a split cascade scenario \citep{Alexakis2018}. In this scenario, from the intermediate scales at which energy is injected, part of the energy is cascaded to small scales as in regular 3D turbulence, while another fraction of the energy is cascaded from the intermediate scales to larger scales, owing to the quasi-2-dimensionalization. This gives rise to a net forward flux of energy at smaller scales, but a net inverse flux of energy at larger scales. While the forward flux is balanced by regular viscous dissipation, we shall here balance the inverse flux of energy with hypoviscosity at the large scales, to suppress the formation of the LSV.

\begin{figure}
    \centering
    \includegraphics[width=0.9\linewidth]{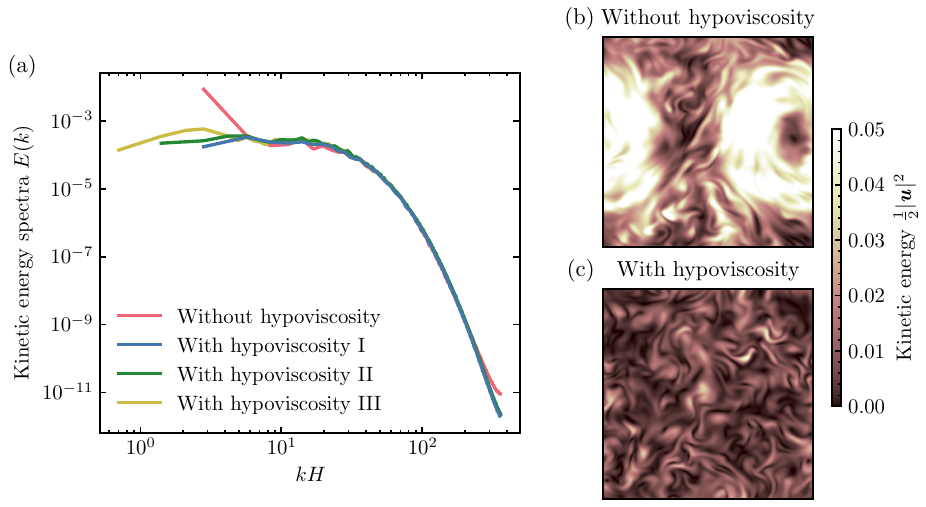}
    \caption{(a) Instantaneous kinetic energy spectra $E(k)$ for the cases with $\textrm{Ra}=8\times10^6$ with and without hypoviscosity and for different horizontal domain sizes (see table~\ref{tab:input}). (b,c) The corresponding snapshots of kinetic energy $\tfrac{1}{2}|\bm{u}|^2$ for the cases with $L/H=2.235$ at the mid-plane $z=H/2$.}
    \label{fig:hypo_example}
\end{figure}

To show the influence of the hypoviscosity on the flow we consider the kinetic energy spectrum
\begin{equation}
    E(k) \equiv \frac{L}{2\pi} \int_0^H \sum_{k\leq \sqrt{k_x^2+k_y^2} < k+2\pi/L} \tfrac{1}{2} \left| \hat{\bm{u}}\left(k_x,k_y,z\right) \right|^2 \mathrm{d}z.
\end{equation}
This is shown for an exemplary case in figure~\ref{fig:hypo_example}. Indeed, we notice that the largest scales (smallest wavenumbers), in which the LSV resides, are strongly damped by the presence of the hypoviscosity, whereas the intermediate and small scales remain largely unaffected.

To tune the hypoviscosity exponent $\alpha$ and strength $\nu_\alpha$ we need to ensure that two conditions are met for all runs in each series:
\begin{enumerate}
    \item The hypoviscosity needs to be able to dissipate the full inverse energy flux before it condenses into an LSV. In practice, this means that the hypoviscosity must be strong enough such that the energy contained in the smallest (system scale) wavenumber does not dominate the spectrum. This can be translated into a condition on the spectral slope at the smallest wavenumber, demanding that \mbox{$\frac{\mathrm{d}\log E(k)}{\mathrm{d}\log k} \big|_{k=2\pi/L} > 0$}.
    \item The action of the hypoviscosity needs to remain restricted to the smallest wavenumbers. We ensure this by checking that the spectral slope at small wavenumbers does not exceed \mbox{$\frac{\mathrm{d}\log E(k)}{\mathrm{d}\log k} \big|_{k=2\pi/L} < 2\alpha$}, such that the action of the hypoviscosity is indeed dominated by the smallest wavenumbers.
\end{enumerate}

Then, if there is sufficient scale separation between the large scales where the hypoviscosity acts and the scales that dominate the energy injection, conservation of energy ensures that the rate of energy dissipation by the hypoviscosity will match exactly with the inverse energy flux, independent of the precise values of the hypoviscosity parameters, as long as the aforementioned conditions are met. This is verified numerically for an exemplary case in appendix~\ref{appA}.

\begin{figure}
    \centering
    \includegraphics[width=0.92\linewidth]{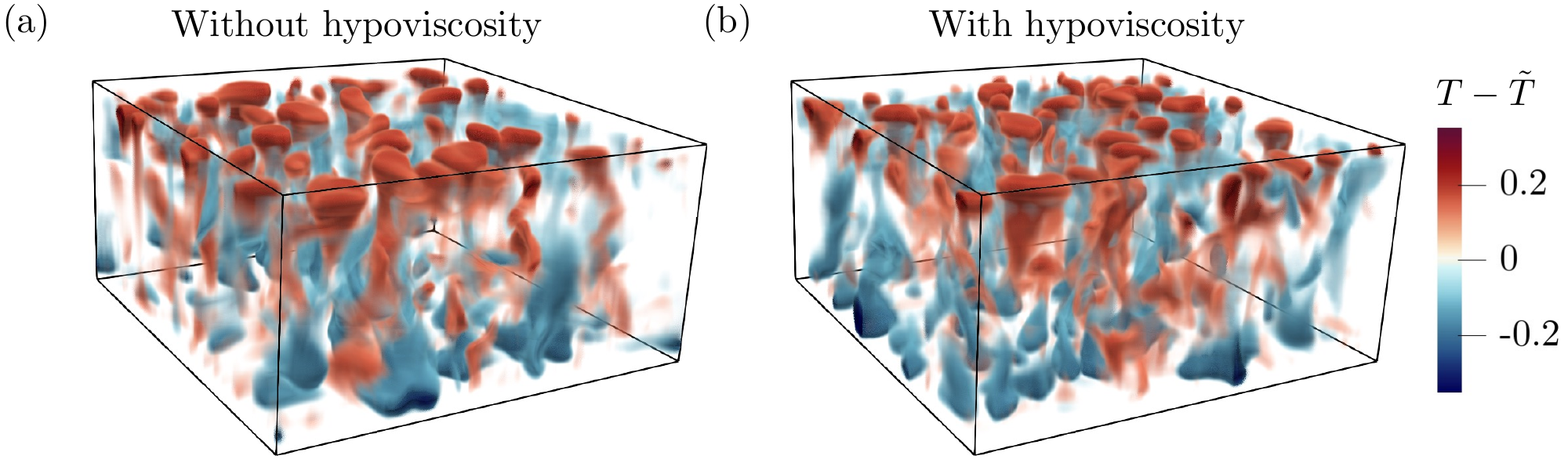}
    \caption{Instantaneous snapshots of the temperature fluctuation field $T-\tilde{T}(z)$, where $\tilde{T}(z)$ is the horizontally and temporally averaged temperature profile, for the cases with $\textrm{Ra}=8\times10^6$ and $L/H=2.235$ without hypoviscosity (a) and with hypoviscosity (b).}
    \label{fig:hypo_render}
\end{figure}

Under these conditions, fluctuation fields that are dominated by intermediate- and small-scale structures remain unaffected by the presence of hypoviscosity, see for example figure~\ref{fig:hypo_render}.

\section{Transition towards inverse cascade} \label{sec:transition} 
With the hypoviscosity in place and tuned, we are well-equipped to study the transition towards the inverse energy cascade under the influence of large-scale friction, in absence of the LSV. To that extent, we consider the strength of the inverse energy flux $\varepsilon_\textrm{inv}$ as we vary the control parameter $\textrm{Ra}$ throughout the regime where the inverse cascade is known to occur.

As argued above, when the action of the hypoviscosity is correctly tuned, the energy dissipated by the hypoviscosity coincides with the inverse energy flux. We can thus measure the mean inverse energy flux in this case as
\begin{equation}
    \varepsilon_\textrm{inv}^\textrm{(hypo)} \equiv \left\langle \nu_\alpha \bm{u}_h \cdot \Delta_h^{-\alpha} \bm{u}_h \right\rangle,
\end{equation}
where we use $\langle ... \rangle$ to denote the average over the full spatial domain and time.

For comparison, we compute the inverse flux for the cases without hypoviscosity directly from the triadic non-linear interactions that transfer kinetic energy from all scales into the system scale 2D mode in which the LSV resides \citep{Verma2019}. The inverse flux is then computed as
\begin{equation}
    \varepsilon_\textrm{inv}^\textrm{(no-hypo)}  \equiv  \sum_Q T_\textrm{3D} (K=\tfrac{2\pi}{L},Q) + T_\textrm{2D} (K=\tfrac{2\pi}{L},Q)
\end{equation}
where, following \citet{Rubio2014,Guzman2020,DeWit2022}, we distinguish baroclinic contributions
\begin{subequations}
\begin{equation}
    T_\textrm{3D}(K,Q)\equiv-\left\langle\bm{u}^\textrm{2D}_K\cdot(\overbar{\bm{u}^\textrm{3D} \cdot \bm{\nabla} \bm{u}^\textrm{3D}_Q})\right\rangle,
\end{equation}
and barotropic contributions
\begin{equation}
    T_\textrm{2D}(K,Q)\equiv-\left\langle\bm{u}^\textrm{2D}_K\cdot(\bm{u}^\textrm{2D}\cdot\bm{\nabla}\bm{u}^\textrm{2D}_Q)\right\rangle,
\end{equation}
\end{subequations}
where the overbar $\overbar{\vphantom{a}...}$ denotes vertical averaging and with $\bm{u}^\textrm{2D} \equiv \overbar{\bm{u}_h}$ and $\bm{u}^\textrm{3D}\equiv\bm{u}-\bm{u}^\textrm{2D}$, and with subscript $K$ denoting Fourier-filtering at mode $K$.

\begin{figure}
    \centering
    \includegraphics[width=\linewidth]{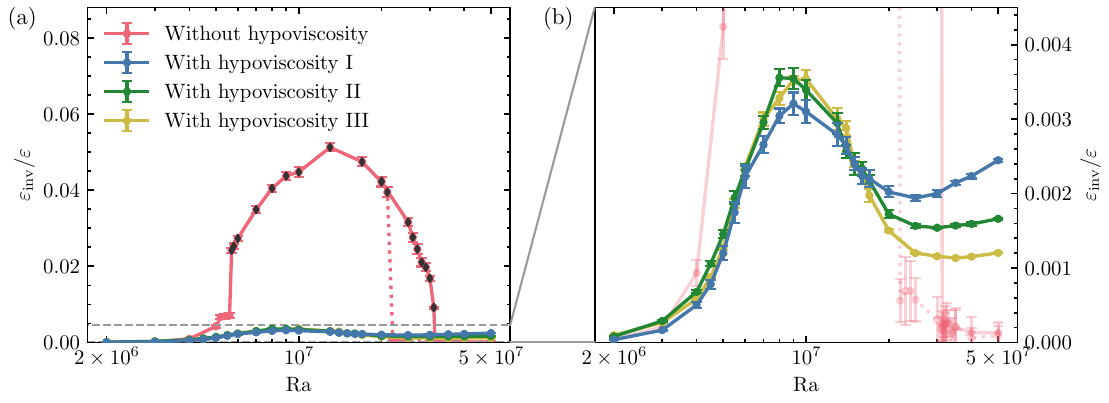}
    \caption{The total inverse flux of kinetic energy $\varepsilon_\textrm{inv}$ normalized by the total energy flux $\varepsilon = (\textrm{Nu}-1)/\sqrt{\textrm{Pr}\textrm{Ra}}$ as a function of $\textrm{Ra}$ for $\textrm{Ek}=10^{-4}$ and $\textrm{Pr}=1$ for the different series of runs with and without hypoviscosity. Panel (b) is a zoom of panel (a) that focuses on the cases with hypoviscosity, indicated by the dashed box in (a). For the runs without hypoviscosity, black diamonds indicate cases that show LSV formation. The figure reveals that while the transition to the inverse cascade without hypoviscosity is discontinuous and hysteretic (lower hysteretic branch is depicted by the dotted line), the transition for the case with hypoviscosity is continuous.}
    \label{fig:inv_series}
\end{figure}

The results are shown in figure~\ref{fig:inv_series}. While the transition to the inverse cascade regime without large-scale friction, in presence of the LSV, is known to be discontinuous and hysteretic \citep{Favier2019,DeWit2022}, we find that this behaviour is entirely different for the case with large-scale friction, in absence of the LSV. Here, we observe that the transition in this case is in fact smooth and the discontinuity disappears (and thereby also the hysteretic behaviour, see appendix~\ref{appC}). This is in agreement with what is found in other hypoviscous quasi-2D systems \citep{Seshasayanan2014,Benavides2017,Pestana2019,VanKan2020}. We furthermore find that the strength of the inverse cascade in absence of the LSV is much smaller than for the cases that have an LSV.

A point of attention is raised by the observation that the inverse cascade does not seem to decay to zero for large \textrm{Ra} for the case with hypoviscosity. However, we argue that this is an effect of insufficient scale separation between the large scales at which the hypoviscosity acts and the scales that dominate the energy injection, such that the hypodissipation not only measures the inverse energy flux, but also the fraction of energy that is directly injected at these large scales. The energy injection is known to be dominated by features of typical size $\mathcal{L}_c$, but the self-organizing nature of the forcing injects energy at a broad range of scales around this dominant scale. The runs without hypoviscosity (red line) were performed at $L\approx 10\mathcal{L}_c$ and the runs with hypoviscosity I (blue line) employ the same aspect ratio in order to have a one-to-one comparison. However, to increase the scale separation, we have added the series of runs in the wider domains $L \approx 20 \mathcal{L}_c$ (II, green line) and  $L \approx 40 \mathcal{L}_c$ (III, yellow line), which also allows for a less aggressive exponent of the hypoviscosity (see table~\ref{tab:input}). Indeed, we notice that the measured energy flux that is dissipated by the hypoviscosity remains mostly unchanged, but reaches a smaller value at large $\textrm{Ra}$ for the wider domains. We thus expect that the inverse flux will decay to values close to zero in the limit of increasingly large domain size as the scales become more cleanly separated.

\begin{figure}
    \centering
    \includegraphics[width=0.8\linewidth]{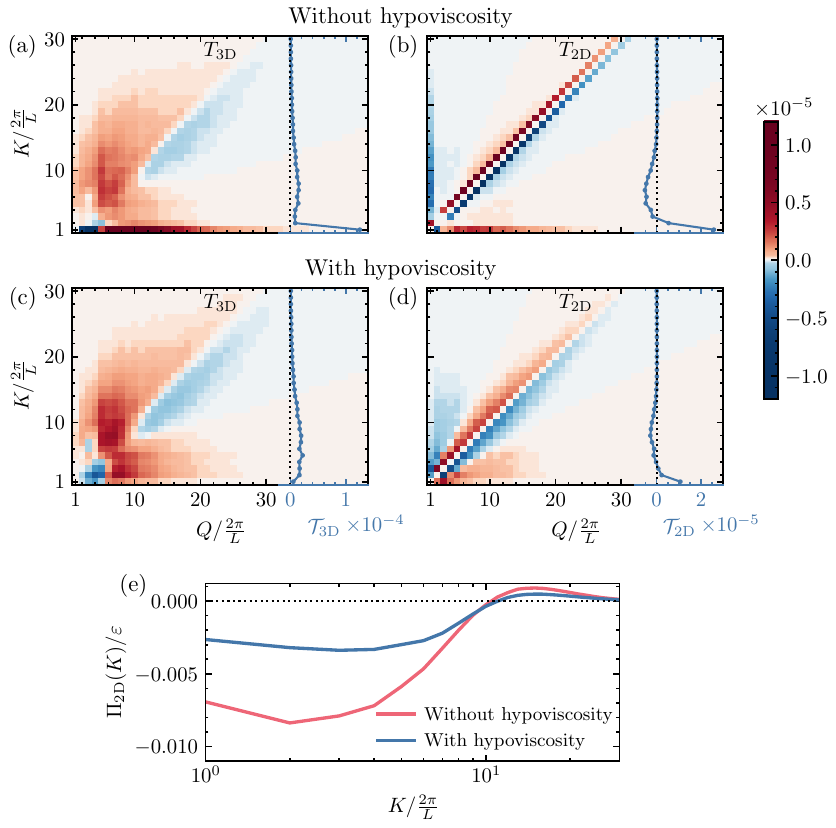}
    \caption{Time-averaged kinetic energy transport maps from 3D (a,c) and 2D (b,d) modes $Q$ to 2D modes $K$, i.e. $T_\mathrm{3D}(K, Q)$ and $T_\mathrm{2D}(K, Q)$, respectively, for the cases with \mbox{$\textrm{Ra}=8\times10^6$} and $L/H=2.235$ without hypoviscosity (a,b) and with hypoviscosity (c,d). Panels (a-d) also show the respective sums over the donating scales $Q$, obtained as \mbox{$\mathcal{T}_\mathrm{3D}(K)\equiv \sum_Q T_\mathrm{3D}(K,Q)$} and \mbox{$\mathcal{T}_\mathrm{2D}(K) \equiv \sum_Q T_\mathrm{2D}(K,Q)$} (blue lines). Panel (e) depicts the total 2D energy flux \mbox{$\Pi_\textrm{2D}(K) \equiv -\sum_{K'<K} \mathcal{T}_\textrm{2D}(K)$}, normalized by the total energy flux $\varepsilon$.}
    \label{fig:trans_maps}
\end{figure}

The findings regarding the inverse cascade transition under the influence of hypoviscosity not only teach us something about what happens to convective systems that are subject to a large-scale friction mechanism, but through comparison, they also teach us about the influence of the LSV in the case without large scale friction. Indeed, since the hypoviscosity only suppresses the LSV at large scales, leaving the other scales of the flow unaffected, we understand that the discontinuity observed in the inverse cascade transition in the case without large-scale friction, is in fact a consequence of interactions between the LSV and the background turbulence.

To further understand the influence of the presence/absence of the LSV on the nature of the inverse cascade, we also consider the scale-to-scale kinetic energy transfer maps. By looking at $T_\textrm{3D}(K,Q)$ and $T_\textrm{2D}(K,Q)$, we can study how baroclinic and barotropic kinetic energy is transferred from donating wavenumber $Q$ to receiving wavenumber $K$. The results are provided in figure~\ref{fig:trans_maps} for an exemplary case with and without hypoviscosity. It shows that in presence of the LSV (upper panels), the upscale energy transfer is highly non-local, with a direct transfer from all wavenumbers $Q$ into the largest wavenumber $K=2\pi/L$, as is also known from \citet{Rubio2014,Guzman2020,DeWit2022}. In absence of the LSV (lower panels), on the contrary, we find that the upscale energy transfer has a much more local signature, with the energy transfer concentrating more around the diagonal of the map, indicating transfer between more like-sized scales. Note furthermore in figure~\ref{fig:trans_maps}e, that the total 2D energy flux $\Pi_\textrm{2D}(K)$ towards low $K$ for the case with hypoviscosity is found to be $\approx 0.003 \varepsilon$, comparable to the total inverse energy flux measured by the hypodissipation (see figure~\ref{fig:inv_series}b). This indicates that the inverse cascade that develops in the 2D manifold is fully dominating the measured total inverse flux in the case with hypoviscosity, while for the case without hypoviscosity (with LSV), the direct (non-local) baroclinic interactions between the large-scale 2D mode and the 3D modes are much stronger contributors (see figure~\ref{fig:trans_maps}a).

\begin{figure}
    \centering
    \includegraphics[width=\linewidth]{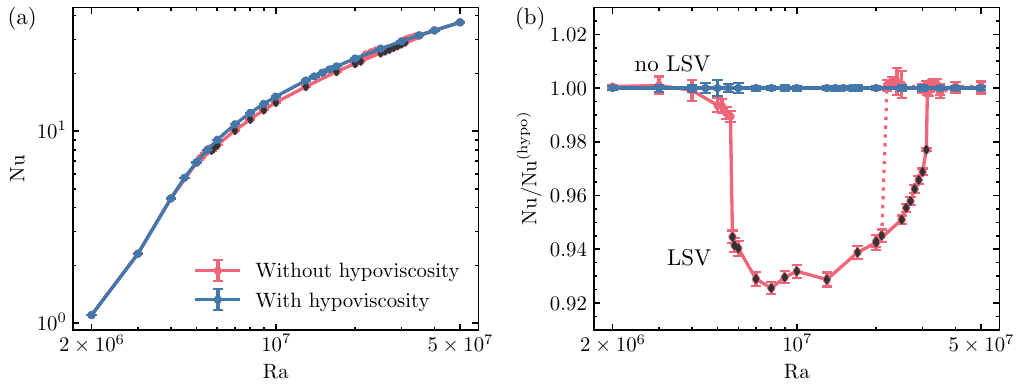}
    \caption{(a) The average Nusselt number $\textrm{Nu}$ as a function of $\textrm{Ra}$ for $\textrm{Ek}=10^{-4}$ and $\textrm{Pr}=1$ for the series of runs with and without hypoviscosity and $L/H=2.235$. (b) The same data compensated by the $\textrm{Nu}$ of the runs with hypoviscosity. For the runs without hypoviscosity, black diamonds denote cases that show LSV formation. This indicates that the LSV lowers the $\textrm{Nu}$ by around 3-8\%.}
    \label{fig:nu_series}
\end{figure}

Finally, we can also study what the influence of the LSV is on the total thermal throughput of the convective system, quantified through the Nusselt number $\textrm{Nu}$, given by
\begin{equation}
    \textrm{Nu} \equiv \sqrt{\textrm{Pr}\textrm{Ra}}\langle T\bm{u}\cdot\bm{e}_z \rangle + 1.
\end{equation}
These results are depicted in figure~\ref{fig:nu_series}. As first observed by \citet{Guervilly2014}, when the LSV emerges, the heat transfer slightly decreases. Without hypoviscosity, however, the $\textrm{Nu}$ for the case without LSV can only be measured in the transient phase before the LSV emerges, at contrast with the statistically steady state obtained here. Comparing the runs with and without hypoviscosity, we find that the suppression of $\textrm{Nu}$ by the LSV remains in the order of 3-8\% throughout the full range of $\textrm{Ra}$ in which the LSV is active for the $\textrm{Ek}=10^{-4}$ considered here. In the cases where there is no LSV, there is a good agreement between the runs with and without hypoviscosity.

It should be noted that the current simulations are conducted with stress-free boundary conditions. When considering no-slip boundaries, the effect of Ekman pumping is known to increase $\textrm{Nu}$ \citep{Stellmach2014}, which could counteract the suppressive effect of the LSV.

\citet{Oliver2023} have studied the scaling of $\textrm{Nu}$ in absence of the LSV in the limit of $\textrm{Ek}\to 0$ using asymptotically reduced equations. Remarkably, they find that in that case, the $\textrm{Nu}$ approaches the diffusivity-free scaling $\textrm{Nu}\sim \textrm{Pr}^{-1/2} \textrm{Ra}^{3/2} \textrm{Ek}^2$. In the current work, the $\textrm{Ek}$ is not sufficiently low to retrieve such scaling.

\section{Conclusions and outlook} \label{sec:conclusion} 
We have shown how a modified form of hypoviscosity can be applied to rotating Rayleigh-B\'{e}nard convection to effectively suppress the formation of a large-scale vortex in the inverse cascade regime. Using this hypoviscosity, we find that the transition towards the inverse cascade is gradual and continuous, at contrast with the discontinuous transition that is found when the LSV is allowed to form. This is a strong indication that the discontinuity is in fact not an intrinsic characteristic of the development of the inverse cascade itself, but is purely a consequence of the interaction between the strong LSV and the background turbulence. This supports the idea of a positive feedback mechanism, where the emergence of the LSV enhances the upscale transport itself, as speculated in \citet{Rubio2014,Favier2019,DeWit2022}. The details of how such a mechanism can arise warrant further investigation, which may extend more generally to other turbulent flow systems with condensate-background interaction.

The continuous transition towards the inverse cascade that is found in this work is in agreement with what is found in other quasi-2D systems, such as magneto-hydrodynamics \citep{Seshasayanan2014}, thin layer turbulence \citep{Benavides2017} and forced rotating turbulence \citep{Pestana2019,VanKan2020}. There, indications for criticality of this transition were found in the limit of increasingly large domain size and high Reynolds number. For rotating convection, this will be hard to prove numerically, given the imperfect scale separation that is inherent to the natural convective forcing. Nonetheless, one may speculate that criticality could manifest in a similar fashion in rotating convection, considering the found phenomenological and theoretical similarities between rotating convection and the other quasi-2D systems.

In a more general sense, the findings in this work fit within a broader class of research that investigates how large-scale structures can non-trivially affect the underlying smaller scale turbulence \citep{Biferale2001,Marusic2017,Berghout2021,Lalescu2021,Alcayaga2022}. The examples of significant interactions between large-scale structures and small-scale turbulence in these works are indications of when the classical view of scale separation and universality of small-scale turbulence starts to become more subtle.

\begin{bmhead}[Acknowledgments.]
The author thanks R.P.J. Kunnen, H.J.H. Clercx and A. Alexakis for useful discussions.
\end{bmhead}

\begin{bmhead}[Funding.]
This work is supported by the Netherlands Organization for Scientific Research (NWO) through the use of supercomputer facilities (Snellius) under Grant No. 2023.026, EINF-11692, EINF-12328. This publication is part of the project “Shaping turbulence with smart particles” with Project No. OCENW.GROOT.2019.031 of the research program Open Competitie ENW XL which is (partly) financed by the Dutch Research Council (NWO).
\end{bmhead}

\begin{bmhead}[Declaration of Interests.]
The author reports no conflict of interest.
\end{bmhead}


\begin{appen}

\section{Other methods of LSV suppression}\label{appB}

In this work, an adapted form of hypoviscosity is used to suppress the formation of the LSV. In practice, however, different mechanisms could be responsible for the suppression of the LSV. A very natural mechanism could be the friction caused by no-slip boundaries, which give rise to Ekman boundary layers that can partially or fully suppress the formation of the LSV \citep{kunnen2016transition,julien2016nonlinear,plumley2016effects,Guzman2020}. Another approach for suppression of the LSV is put forward by \citet{Oliver2023}, where the non-linear term of the Navier-Stokes equations is modified to remove the non-linear interactions in the 2D vertically averaged manifold, modifying equation \eqref{eq:NS_momentum} as
\begin{equation}
\frac{\partial\bm{u}}{\partial t} + \left(\bm{u}\cdot\bm{\nabla}\right)\bm{u}  - \overbar{\left(\bm{u}\cdot\bm{\nabla}\right)\bm{u}}+ \frac{\left(\textrm{Pr}/\textrm{Ra}\right)^{1/2}}{\textrm{Ek}}\bm{e}_z\times\bm{u} = -\bm{\nabla}p + T\bm{e}_z + \left(\frac{\textrm{Pr}}{\textrm{Ra}}\right)^{1/2}\Delta \bm{u} .
\end{equation}

In figure \ref{fig:lsv_suppression_comp}, we compare the different methods of LSV suppression for an exemplary case. It shows that for the case considered here, all mechanisms yield a complete suppression of the LSV formation at large scales, while other scales remain mostly unaffected. The simulation with hypoviscosity yields the least aggressive suppression of the large scale energy. We finally point out that in our work, the hypoviscosity not only serves to suppress the formation of the LSV, but also serves as a tool to measure the strength of the inverse cascade. How to harness other mechanisms in order to measure the strength of the inverse cascade may be less straightforward.

\begin{figure}[htb]
    \centering
    \includegraphics[width=\linewidth]{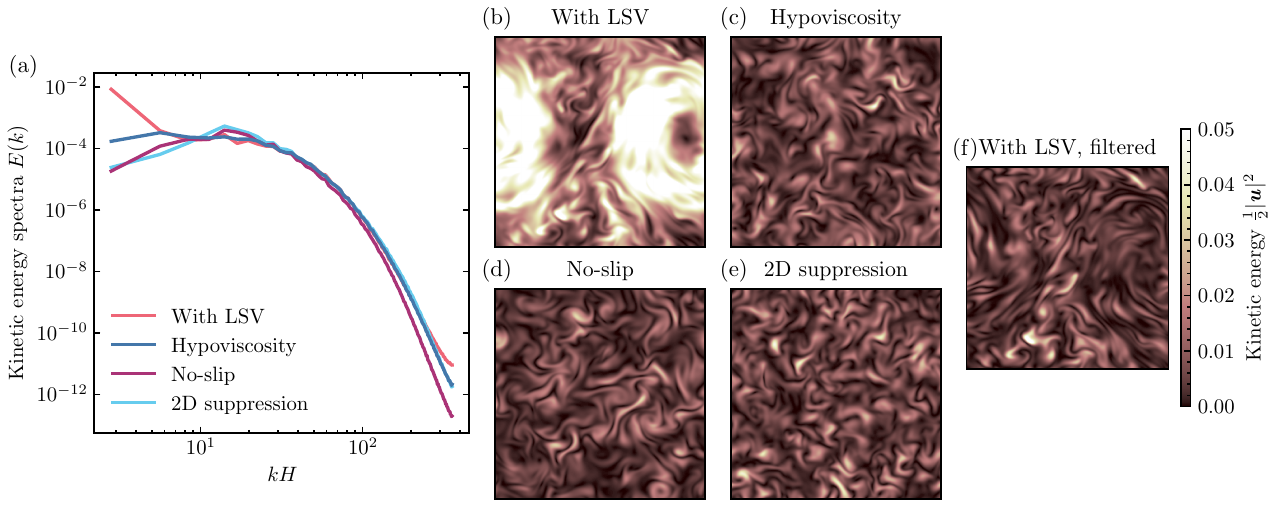}
    \caption{Comparison of different methods of LSV suppression for an exemplary case with $\textrm{Ra}=8\times10^6$ and $L/H=2.235$, depicted through instantaneous kinetic energy spectra (a) and snapshots of kinetic energy at the mid-plane (b-e). We consider the case with LSV, without suppression (b), the case with hypoviscosity as treated in the main text (c), the case with no-slip top and bottom wall boundary conditions (d) and the case with suppression of 2D non-linear interactions (e). For comparison, we also show the case with LSV where the $k=2\pi/L$ modes have been fully filtered out \textit{a posteriori} (f).}
    \label{fig:lsv_suppression_comp}
\end{figure}

\section{Validation with varying hypoviscosity}\label{appA}

When the large scales at which the hypoviscosity is active are sufficiently well separated from the intermediate scales that dominate the energy injection, the rate at which the hypoviscosity dissipates kinetic energy becomes precisely equal to the inverse flux of kinetic energy, independent of the exact parameters of the hypodissipation. This is analogous to the relation between energy injection rate and viscous dissipation rate in regular 3D turbulence.

We show this for an exemplary case in figure~\ref{fig:appendix}. Indeed, for the run with higher hypoviscosity coefficient $\nu_\alpha=0.076$, the spectral energy at the large scales naturally becomes smaller, such that the total hypodissipation becomes equal to the run with smaller hypoviscosity coefficient $\nu_\alpha=0.038$, with both agreeing within error bars. This confirms that here, the hypodissipation is indeed well tuned and it is a good proxy for the total inverse flux.

\begin{figure}[h!]
    \centering
    \includegraphics[width=\linewidth]{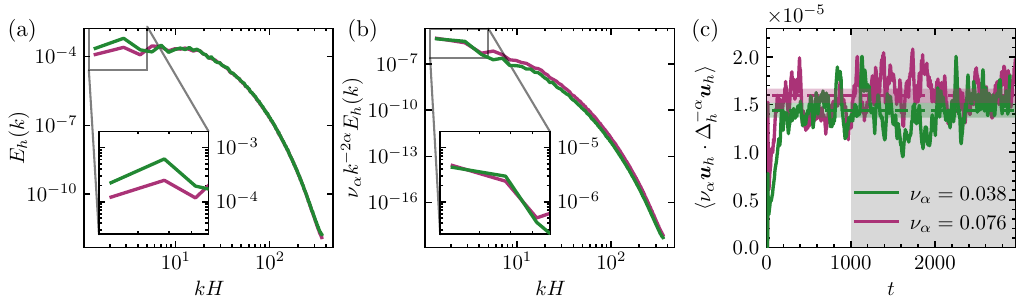}
    \caption{Comparison of runs at $\textrm{Ra}=8\times10^6$ and $L/H=4.470$ with two different values of the hypoviscosity coefficient $\nu_\alpha=0.038$ (green lines) and $\nu_\alpha=0.076$ (yellow lines) while both have hypoviscosity exponent $\alpha=1$. Depicted through the instantaneous kinetic energy spectra of horizontal velocity $E_h(k)$ (a) and hypodissipation spectra $\nu_\alpha k^{-2\alpha} E_h(k)$ (b) as well as timeseries of the total hypodissipation (c). In panel (c) the gray shaded area indicates the time interval in which the statistically steady state is reached that is used for averaging, while the horizontal dashed lines indicate the averages themselves, with statistical errorbars as coloured shaded areas.}
    \label{fig:appendix}
\end{figure}

\section{Validation of absence of hysteresis}\label{appC}

In earlier works, it has been observed that the transition into the LSV state at high $\textrm{Ra}$ is discontinuous and hysteretic \citep{Favier2019,DeWit2022}. In the current work, we show that when the LSV is suppressed using a friction mechanism at large scales, the transition towards the inverse cascade regime becomes continuous. As a result, we expect that also the hysteresis in the transition should vanish. In figure~\ref{fig:hypo_hyst}, we test this explicitly for an exemplary case with $\textrm{Ra}=2.5\times 10^7$, which is in the range where hysteresis was observed for the case with LSV \citep{DeWit2022}. Indeed, comparing the run that is freshly initialized as described in the main text, to the runs that are restarted from a lower $\textrm{Ra}=1.0\times 10^7$ and a higher $\textrm{Ra}=4.0\times 10^7$, we confirm that there is no statistical dependence on the initial conditions.

\begin{figure}[htb]
    \centering
    \includegraphics[width=0.6\linewidth]{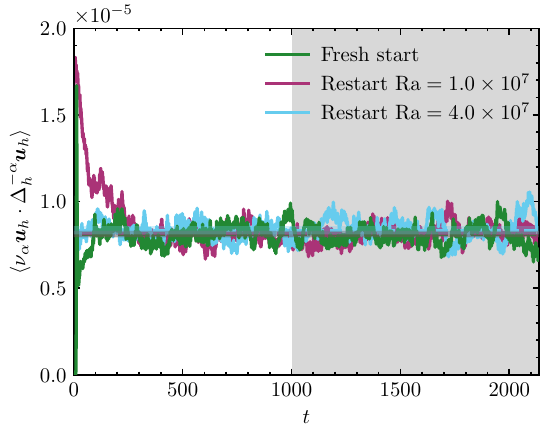}
    \caption{Timeseries of hypodissipation for the case with $\textrm{Ra}=1.5\times10^7$ and $L/H=4.470$ for a run that is freshly initialized (green line), a run that is restarted from a run with $\textrm{Ra}=1.0\times 10^7$ (purple line) and a run that is restarted from a run with $\textrm{Ra}=4.0\times 10^7$ (light blue line). The gray shaded area indicates the time interval in which the statistically steady state is reached that is used for averaging, while the horizontal dashed lines indicate the averages themselves, with statistical errorbars as coloured shaded areas.}
    \label{fig:hypo_hyst}
\end{figure}

\end{appen}

\bibliographystyle{jfm}
\bibliography{main}

\end{document}